\documentstyle[12pt]{article}
\begin{document}
\centerline{\bf Right-Handed b to c and u Coupling Model and CP Violation }
\baselineskip=8truemm
\vskip 2.0truecm
\centerline{Toshio HAYASHI  \footnote{e-mail: hayashi@ed.
kagawa-u.ac.jp} }
\vskip 0.8truecm
\centerline{\it  Department of Physics, University of Kagawa, 
Takamatsu 760, JAPAN}
\vskip 2.0truecm

\centerline{\it ABSTRACT}

\vskip 0.5truecm

We investigate CP violation in the purely right-handed b quark 
to c- and u- quark coupling model under the constraint of right-handed
W-gauge boson mass $M_R>720$ GeV , which is experimentally obtained 
recently by 
D0 Collaboration at Fermilab.\  By using the data on $K_L-K_S$ mass 
difference, CP violating parameter $\varepsilon$ in the neutral kaon system
and $B_d-\bar B_d$ mixing, together with the new data of ${\rm Br}
(B^-\rightarrow \psi\pi^-) /{\rm Br}(B^-\rightarrow \psi K^-) = 
0.052 \pm 0.024~ (\approx~
\mid V_{cd}/V_{cs} \mid^2)$, we can fix all of the three independent angles
and one phase of the right-handed mixing matrix $V^R$. Under these
constraints, another CP-violating parameter $\varepsilon'/\varepsilon$
and electric dipole moment of neutron are shown to be consistent with
the data in our model. The pattern of CP violation in the nonleptonic
decay of $B_d(B_s)$ mesons to CP eigenstates is different from that in the
Standard Model.

\newpage
\leftline{\bf 1.\ \ Introduction}
\vskip 0.4truecm
\noindent
The Standard Model of electroweak interactions for quarks and leptons 
has proved to be increasingly successful by the discovery of top 
quark by CDF and D0 Collaborations \cite{Abe94}.

All of the weak charged currents are postulated to be left-handed
(V-A) in the Standard Model.\  As for the quark currents, among the nine
charged currents caused by the Cabibbo-Kobayashi-Maskawa quark
mixing \cite{Kobayashi73}, those between the first two generations and 
the diagonal third generation coupling $t\rightarrow b$ are known to be 
left-handed \cite{Barbiellini86,Sohaile92}.\ 
However, the chirality of the off-diagonal couplings in the $b$- quark
decay, $b\rightarrow c$ and $b\rightarrow u$, is not yet experimentally
confirmed to be left-handed. \ Polarization of $\Lambda_b$ baryon produced 
in $Z^0\rightarrow b\bar{b}$ decay has recently been measured to be 
$P_{\Lambda_b}=-0.23\pm 0.25$ by using the semileptonic decays with
charmed hadrons at LEP \cite{BuskulicI96}, while it has been predicted to be 
$P_{\Lambda_b}\simeq -0.75$ in a simple model in the framework of the 
Standard Model.

Since the $SU(2)_L\times SU(2)_R \times U(1)$ gauge model with right-handed 
charged currents was proposed \cite{Pati74}, the mass $(M_R)$ of right-handed
gauge boson $W_R$ has been studied phenomenologically.\  Beal, Bander and 
Soni obtained the constraint $M_R>1.6$ TeV from the analysis of $K_L-K_S$ 
mass difference
in the left-right symmetric model $(V^R=V^L)$ \cite{Beall82}, 
where $V^L$ and $V^R$ are left- and right-handed quark mixing matrices, 
respectively.\  After that, Olness and Ebel discovered two types of 
$V^R$ which 
enabled to lower the mass $M_R$ to several-hundred GeV by relaxing the 
constraint $V^R=V^L$ \cite{Olness84}. \ Langacker and Sanker extended 
their analysis and showed that the mass can be lowered to 300GeV for 
the case of right-handed neutrino mass of above 100 MeV for the special 
form of $V^R$ \cite{Langacker89} which is idealized from the types found 
by Olness and Ebel. Nishiura, Takasugi and Tanaka proposed a model with 
light $W_R$ in which CP violation in $K\rightarrow \pi \pi$ decay 
mainly arises 
from the right-handed interactions, and investigated CP violation in 
$K$ and $B_{d(s)}$ decays and the electric dipole moment of neutron 
\cite{Nishiura90}.

In 1992, Gronau and Wakaizumi proposed a purely right-handed $b$ to $c(u)$
coupling model \cite{Wakaizumi92} in the framework of 
$SU(2)_L\times SU(2)_R\times U(1)$
gauge model under the circumstance that the left-handed chirality of these
couplings has not yet been confirmed experimentally, and showed that 
this model is viable within both the experimental precision of $b$ semileptonic
decays, $K_L-K_S$ mass difference and $B_d-\bar{B_d}$ mixing and 
the theoretical uncertainties in the hadronic matrix elements.\ After that,
Hou and Wyler discussed these couplings more generally in their model 
\cite{Hou92}. \ Several methods to test the chirality of the $b$ to $c$ 
coupling have been proposed \cite{Rosner93}.

As for CP violation in the purely right-handed $b$ to $c$ coupling model,
Gronau investigated CP asymmetry in neutral $B$ meson decays into 
hadronic CP-eigenstates by giving one phase to $V^R$ and obtained a
remarkably different pattern of the asymmetry from that in the Standard
Model \cite{Gronau92}. \ Hattori, Hasuike, Hayashi and Wakaizumi proposed a 
model of purely right-handed $b$ to $c$ coupling together with purely 
left-handed $b$ to $u$ one with a full number of phases in $V^R$ in order to 
examine CP violation in neutral kaon system, in neutral $B$ meson decays and
electric dipole moment of neutron, and discussed the behavior of the
phases \cite{Hattori95}.

A new lower bound of the right-handed gauge boson $W_R$ has recently been
obtained to be 720 GeV by D0 Collaboration at Fermilab \cite{Abachi96}.\  
In this paper, under this new circumstance we investigate  the model of purely
righted-handed chirality for both $b$ to $c$ and $b$ to $u$ couplings
by giving the most general form to $V^R$, that is, with the three independent
angles and all of the necessary phases to study CP violation in neutral
$K$ and $B$ systems  and the electric dipole moment of neutron. 
  
In $\S$2, we describe the model and determine some of the parameters in the 
model by using the data on $K_L-K_S$ mass difference, CP-violating parameter
$\varepsilon$ in the neutral kaon system, $B_d-\bar{B}_d$ mixing and the new
data on ${\rm Br}(B^-\rightarrow \psi \pi^-)/{\rm Br}(B^-\rightarrow \psi K^-)
=0.052\pm 0.024 (\approx|V_{cd}/V_{cs}|^2)$. In $\S$3, we examine another 
CP violating parameter
$\varepsilon'$, electric dipole moment of neutron and CP asymmetries in
$B_{d(s)}$ meson decays into hadronic CP-eigenstates. We present discussions 
and conclusions in $\S$4.    
\vskip 1truecm
\leftline{\bf 2.\  Purely right-handed b to c(u) model } 

\vskip 0.4truecm
\noindent
Our model is the purely right-handed coupling model for both $b$- quark
to $c$ and $u$ charged currents in the framework of $SU(2)_L \times 
SU(2)_R \times U(1)$ gauge model, so that the left-handed quark mixing 
matrix is
\begin{equation}
    V^L \simeq
    \left( \begin{array}{ccc}
            1      & \lambda &   0   \\
          -\lambda &   1     &   0   \\
              0    &   0     &   1
           \end{array}
     \right), 
\label{shiki1}
\end{equation}
where the Wolfenstein parametrization with the Cabibbo angle 
$\lambda \equiv \sin \theta_c (\simeq 0.22)$ is used. \ For the right-handed 
mixing matrix, we take the following most general form with all of the 
three mixing angles $\theta_{12}, \theta_{13}\  \mbox{and}\  \theta_{23}$,
\begin{equation}
    V^R =
    \left( \begin{array}{lll}
e^{i\alpha}c_{12}c_{13} & -e^{i\beta}s_{12}c_{13}    & e^{i\gamma}s_{13}  \\
e^{i(\alpha-\gamma)}(s_{12}c_{23}e^{i\delta}-c_{12}s_{13}s_{23})
         &  e^{i(\beta-\gamma)}(c_{12}c_{23}e^{i\delta}+s_{12}s_{13}s_{23})
                                                     & c_{13}s_{23}       \\ 
e^{i(\alpha-\gamma)}(-s_{12}s_{23}e^{i\delta}-c_{12}s_{13}c_{23})
         &  e^{i(\beta-\gamma)}(-c_{12}s_{23}e^{i\delta}+s_{12}s_{13}c_{23})
                                                     & c_{13}c_{23}  
           \end{array}
     \right),
\label{shiki2}
\end{equation}
where $c_{12} \equiv \cos \theta_{12}$,\ $s_{12} \equiv \sin\theta_{12}$,
and etc.\ As for CP phases, our $V^L$ needs three independent phases from
its unitarity up to the overall phase and the most general $V^R$
needs six phases. \ From these nine phases, five phases can be eliminated 
by the quark fields, so that four phases survive
in our model and we will assign them to $V^R$ as seen in Eq.(\ref{shiki2}).

In this model, since $b$ quark decay proceeds through the right-handed
charged current (V+A), semileptonic decay $b \rightarrow cl^-\bar{\nu}$
is mediated by the right-handed gauge boson $W_R$ and the following 
relation must be satisfied \cite{Wakaizumi92},
\begin{equation}
|V^R_{cb}| \sqrt{\beta^2_g + \zeta^2_g}=|V^{SM}_{cb}|, 
\label{shiki3}
\end{equation}
where \(\beta_g \equiv (g_R/g_L)^2 M^2_L/M^2_R \)  and
\( \zeta_g \equiv (g_R/g_L) \zeta \), $g_L$ , $g_R$, and $M_L$, $M_R$
being left- and right-handed gauge coupling constants and gauge boson
masses, respectively, $\zeta$ the $W_L-W_R$ mixing angle, and $V^{SM}_{cb}$
is the (cb)-element of the Kobayashi-Maskawa mixing matrix \cite{Kobayashi73}. 
\ $W_R$ has recently been searched for by D0 Collaboration at Fermilab 
\cite{Abachi96} by decays to an electron and a massive right-handed neutrino
($N_R$), $W^{\pm}_R \rightarrow e^{\pm}N_R$, and a lower limit of the mass has 
been obtained as $M^g_R(\equiv(g_L/g_R)M_R)>720$GeV for $m_{N_R}\ll M_R$.
\ An upper limit of $\zeta_g$ is obtained by the phenomenological analysis
with the data on both low-energy processes and high-energy processes
to be \( |\zeta_g|<0.031 \) for light right-handed neutrino \cite{Polak92}.
\ If we use these limits and \( |V^{SM}_{cb}|=0.032-0.048 \) 
\cite{Neubert91},the relation of Eq.(\ref{shiki3})  leads to
\begin{equation}
|V^R_{cb}|=|c_{13}s_{23}|>0.96.
\label{shiki4}
\end{equation}
The constraint, \( |V^{SM}_{ub}/V^{SM}_{cb}|=0.08\pm 0.02 \) \cite{Barnett96}, 
obtained from the analyses of the lepton energy spectra in $B$ meson
semileptonic decays with various theoretical hadronic models is 
transformed into
\begin{equation}
\left|\frac{V^R_{ub}}{V^R_{cb}} \right|
 =\left| \frac{s_{13}}{c_{13}s_{23}} \right|=0.08 \pm 0.02 
\label{shiki5}
\end{equation}
in our model, since we assume \( m_{{\nu}_R}<m_b-m_c \) , where $m_b$
and $m_c$ are the masses of $b$- and $c$- quark, respectively. \ From 
Eqs.(\ref{shiki4}) and (\ref{shiki5})  , we obtain the following range 
for the angle $\theta_{13}$ of $V^R$,
\begin{equation}
|s_{13}|\simeq 0.08 \pm 0.02 . 
\label{shiki6}
\end{equation}

Next, in order to fix the remaining two angles $\theta_{12}, \theta_{23}$
and the phase $\delta$, we use experimental data on $B_d-\overline B_d$ 
mixing, \( {\rm Br}(B^- \rightarrow \psi \pi^-)/{\rm Br}(B^- \rightarrow 
\psi K^-)=0.052\pm 0.024\) obtained recently by CLEO Collaboration 
\cite{Bishai96}, 
$K_L-K_S$ mass difference and CP violating parameter $\varepsilon$ in 
neutral kaon system.

First of all, $B_d-\overline B_d$ mixing is dominantly described in our
model by the $W_L-W_R$ box diagram with $c$- and $t$- quark exchanges,
depicted in Fig.1. \ This diagram gives the following mass difference
between the two mass-eigenstates in neutral $B_d-\overline B_d$ system,
\begin{eqnarray}
\Delta m^{LR}_{B_d}(c,t)&=& 2|<B_d|H^{LR}_{eff}(c,t)|\bar{B_d}>| \nonumber\\
&=& \frac{4G_F^2 M^2_L \beta_g}{\pi^2}
   \left[ {\left(\frac{m_{Bd}}{m_b+m_d} \right)}^2+\frac16 \right]
   \frac14 f^2_B B_B m_{B_d} \nonumber\\
&\times & \left|V^L_{tb} V^{L*}_{cd} V^{R*}_{td} V^{R}_{cb}\right|
   B(x_c,x_t,\beta_g),
\label{shiki7}
\end{eqnarray}
where $m_{B_d}$, $f_B$ and $B_B$ are the mass, decay constant and bag
parameter of $B_d$ meson, respectively, and $m_d$ the $d$- quark mass.
$B(x_c,x_t,\beta_g)$ is the $W_L-W_R$ box function with QCD corrections
which is formulated by Ecker and Grimus \cite{Ecker85} and extended by 
Nishiura, Takasugi and Tanaka \cite{Nishiura90}, where 
$x_c \equiv(m_c/M_L)^2$ and $x_t \equiv (m_t/M_L)^2$ for $c$- and 
$t$- quark mass $m_c$ and $m_t$.\ 
The QCD parameter $\Lambda_f$ of the strong-coupling constant $\alpha_s(m^2)$
in $B(x_c,x_t,\beta_g)$ is chosen to be $\Lambda_f=0.11$ GeV for $N_f=5$ 
to reconcile with $\alpha_s(M^2_Z)=0.122\pm 0.007$ \cite{Barnett96} obtained 
from the event shape measurements by PEP/PETRA, TRISTAN, LEP, SLC and CLEO.
\ If we substitute $G_F=1.166 \times10^{-5} \mbox{GeV},\ M_L=80\mbox{GeV},\
f_B \sqrt{B_B}=(0.15\pm0.05)\mbox{GeV},\ m_{B_d}=5.28\mbox{GeV},\ 
m_b=4.5\mbox{GeV},\ m_d=0.01\mbox{GeV},\ m_c=1.5\mbox{GeV and}\ m_t=174 $ GeV
in Eq.(\ref{shiki7}) and use the data of \( \Delta m_{B_d}=(3.36 \pm 0.39)
\times 10^{-10}\) MeV \cite{Barnett96}, we obtain 
\begin{equation}
|V^R_{td}|=|e^{i\delta}s_{12}s_{23}+c_{12}c_{23}s_{13}|=0.08-0.41,
\label{shiki8}
\end{equation}
where we used Eq.(4) for $|V^R_{cb}|$. \ As for $\beta_g$, we will take 
$M^g_R=750$ GeV to be compatible with the experimental limit
$M^g_R>$720 GeV \cite{Abachi96} hereafter in this paper.

Next, we obtain from the data on the ratio of two nonleptonic decay 
branching ratios $R\equiv {\rm Br}(B^- \rightarrow \psi \pi^-)/{\rm Br}
(B^- \rightarrow \psi K^-)$  \cite{Bishai96} the following constraint 
on the angles, 
\begin{equation}
R\approx { \left| \frac{V^R_{cd}}{V^R_{cs}}\right| }^2
=\frac{ {\left|{ s_{12}c_{23}e^{i\delta}-c_{12}s_{13}s_{23}} \right| }^2}
 {{\left|{c_{12}c_{23}e^{i\delta}+s_{12}s_{13}s_{23}} \right|}^2 }
=0.052\pm0.024 .
\label{shiki9}
\end{equation}

As for $K_L-K_S$ mass difference $\Delta m_K$, three box diagrams depicted in 
Fig.2 contribute in our model. \ The diagram of Fig.2(a) is the same as the
one in the Standard Model and gives the following expression to $\Delta m_K$.
\begin{eqnarray}
\Delta m^{LL}_K(c,c) &=& 2 {\rm Re} <K^0|H^{LL}_{eff}(c,c)|\overline K^0> 
\nonumber\\
  &=& \frac{G_F^2 M^2_L}{6\pi^2} f^2_K B_K m_K {\rm Re}[(V^L_{cs} 
V^{L*}_{cd})^2] \eta^{LL}_{cc}S(x_c), 
\label{shiki10}
\end{eqnarray}
where $m_K$ $f_K$ and $B_K$ are the mass, decay constant and bag parameter
of the kaon, respectively, $\eta^{LL}_{cc}(\sim0.7)$ the QCD correction factor,
and $S(x_c)$ is the Inami-Lim box function \cite{Inami81}. \ If we put 
$f_K=0.16$ GeV and the theoretical range $B_K= \frac13 -1.5$ in 
Eq.(\ref{shiki10}) , we obtain
\begin{equation}
\Delta m^{LL}_K (c,c)=(\frac13-1.5) \times 2.22 \times 10^{-12}{\rm MeV} .
\label{shiki11}
\end{equation}
The diagrams of Figs.2(b) and 2(c) give the following expressions,
\begin{eqnarray}
\Delta m^{LR}_{K}(c,c)&=& \frac{4G_F^2 M^2_L \beta_g}{\pi^2}
   \left[ {\left(\frac{m_{K}}{m_s+m_d} \right)}^2+\frac16 \right]
   \frac14 f^2_K B_K m_K \nonumber\\ 
   &\times& {\rm Re}(V^L_{cs} V^{L*}_{cd} V^{R}_{cs} V^{R*}_{cd})
   B(x_c,x_c,\beta_g),
\label{shiki12}
\end{eqnarray}
\begin{equation}
\Delta m^{RR}_{K}(t,t)= \frac{G_F^2 M^2_L \beta_g}{6 \pi^2}
   f^2_K B_K m_K {\rm Re}[( V^R_{ts} V^{R*}_{td})^2]\eta^{RR}_{tt} S(x^R_t),
\label{shiki13}
\end{equation}
where $x^R_t\equiv (m_t/M^g_R)^2$. \ If we take the $s$- quark mass 
$m_s=0.2$ GeV and QCD correction factor $\eta^{RR}_{tt}\simeq 0.7$ in 
Eqs.(\ref{shiki12}) and (\ref{shiki13}) , the following contributions 
are obtained,
\begin{equation}
\Delta m^{LR}_K(c,c)= (\frac13 -1.5)\times 59.0 \times 
            {\rm Re}(V^R_{cs}V^{R*}_{cd}) \times 10^{-12}\mbox{MeV} ,\\
\label{shiki14}
\end{equation}
\begin{equation}
\Delta m^{RR}_K(t,t)= (\frac13 -1.5)\times 77.8 \times 
            {\rm Re}[(V^R_{ts}V^{R*}_{td})^2] \times 10^{-12}\mbox{MeV} .
\label{shiki15}
\end{equation}
From the requirement that the sum of the three contribution, 
Eqs.(\ref{shiki11}) , (\ref{shiki14}) and (\ref{shiki15}) , gives 
the experimental value of $\Delta m_K$ \cite{Barnett96}, 
we obtain the following constraint,
\begin{equation}
\Delta m_K=\Delta m^{LL}_K(c,c)+\Delta m^{LR}_K(c,c)+\Delta m^{RR}_K(t,t)
=(3.491\pm 0.009)\times 10^{-12}\mbox{MeV}.
\label{shiki16}
\end{equation}

Finally, CP-violating parameter $\varepsilon$ in the neutral kaon system
is given by
\[ \varepsilon = e^{i\pi /4} (\varepsilon_m +\frac1{\sqrt{2}}\, \xi_0), \]
where $\varepsilon_m$ is the contribution from the $K^0-\bar{K}^0$ mixing,
$\varepsilon_m={\rm Im} M_{12}/{\rm Re} M_{12}$, $M_{12}$ being the mass 
matrix of $K^0-\bar{K}^0$ system, and $\xi_0$ is the contribution from decay 
dynamics, $\xi_0={\rm Im} A_0/{\rm Re} A_0$, $A_0$ being the decay amplitude of
$\bar{K}^0 \rightarrow \pi \pi$ for the isospin $I=0$ state of the two
pions. \ As seen in the next section, $\xi_0/\sqrt2$ is about $-0.05$ $\times
10^{-3}$ and constitutes only 2\% of the experimental value of $\varepsilon$
in our model, so that we can neglect the contribution from $\xi_0$ for 
the discussion of $\varepsilon$. \ The contributions to $\varepsilon_m$  come
from the two diagrams of Figs.2(b) and (c) as follows, 

\begin{eqnarray}
\varepsilon^{LR}(c,c)&=& \frac{{\rm Im}<K^0|H^{LR}_{eff}(c,c)|\bar{K}^0>}
                    {\sqrt{2}\,\Delta m_K} \nonumber\\
&=& \frac{2G_F^2 M^2_L \beta_g}{\sqrt{2} \pi^2 \Delta m_K}
   \left[ {\left(\frac{m_{K}}{m_s+m_d} \right)}^2+\frac16 \right]
   \frac14 f^2_K B_K m_K \nonumber\\ 
   &\times& {\rm Im}(V^L_{cs} V^{L*}_{cd} V^{R}_{cs} V^{R*}_{cd})
   B(x_c,x_c,\beta_g),
\label{shiki17}
\end{eqnarray}
\begin{equation}
\varepsilon^{RR}(t,t)= \frac{G_F^2 M^2_L \beta_g}{12\sqrt{2} \pi^2 \Delta m_K}
   f^2_K B_K m_K {\rm Im}[( V^R_{ts} V^{R*}_{td})^2]\,\eta^{RR}_{tt} S(x^R_t).
\label{shiki18}
\end{equation}
When we calculate the two quantities of Eqs.(17) and (18) with the data
$\Delta m_K=(3.491\pm0.009)\times 10^{-12}$ MeV, we obtain
\begin{equation}
\varepsilon^{LR}(c,c)= 5.94 \times {\rm Im}(V^{R}_{cs} V^{R*}_{cd}),\\
\label{shiki19}
\end{equation}
\begin{equation}
\varepsilon^{RR}(t,t)= 7.84 \times {\rm Im}[( V^R_{ts} V^{R*}_{td})^2].
\label{shiki20}
\end{equation}
Therefore, we get the following constraint by summing the contributions
of Eqs.(\ref{shiki19}) and (\ref{shiki20}) and using the data on $\varepsilon$ 
\cite{Barnett96},
\begin{equation}
\varepsilon/e^{i\pi/4}=\varepsilon^{LR}(c,c)+\varepsilon^{RR}(t,t)
=(2.28\pm0.02)\times 10^{-3} .
\label{shiki21}
\end{equation}
This constraint involves three phases $\alpha$, $\beta$ and $\delta$ of
$V^R$ as can be seen in Eq.(\ref{shiki2}). 
The discussion of another CP-violating 
parameter $\varepsilon'$ in the next section will give us a relation
between the two phases, $\alpha-\beta=\pi$. So, the constraint of 
Eq.(\ref{shiki21}) fixes the phase $\delta$.

All of the constraints obtained above, Eqs.(\ref{shiki4}),\,(\ref{shiki6}),
\,(\ref{shiki8}),\,(\ref{shiki9}),\,(\ref{shiki16}) and (\ref{shiki21}),
\,can lead to a solution of the three angles $\theta_{12},
\theta_{13},\theta_{23}$ and one phase $\delta$ of $V^R$,
\begin{eqnarray}
s_{12}&\equiv&\sin\theta_{12}=-0.152,\quad s_{13}=0.09,\quad s_{23}=0.975, 
      \quad \sin\delta=0.0345, \nonumber\\
c_{12}&\equiv&\cos\theta_{12}=0.988,\quad \quad c_{13}=0.996,\quad c_{23}=0.22,
      \quad \cos\delta=-1.0.
\label{shiki22}
\end{eqnarray}
If we substitute the solution of Eq.(\ref{shiki22}) with $\alpha-\beta=\pi$ 
into $V^R$, we obtain the following solution of $V^R$,
\begin{equation}
    V^R =
    \left( \begin{array}{lll}
      e^{i\alpha}0.984      &   -e^{i\alpha}0.151    &  e^{i\gamma}0.09   \\
      e^{i(\alpha-\gamma)}(-0.033e^{i\delta}-0.0867)
                            & -e^{i(\alpha-\gamma)}(0.217e^{i\delta}-0.0133)
                                                           &   0.971      \\
      e^{i(\alpha-\gamma)}(0.148e^{i\delta}-0.0196)
                            & -e^{i(\alpha-\gamma)}(-0.963e^{i\delta}-0.0030)
                                                           &   0.219
           \end{array}
     \right).
\label{shiki23}
\end{equation}
This solution is quite close to the type II of $V^R$ among the four forms
discussed by Langacker and Sanker \cite{Langacker89}, which are 
idealized from the two solutions obtained by Olness and Ebel 
\cite{Olness84} by reconciling the left-right
symmetric model to $K_L-K_S$ mass difference. \ So, the solution of the three
angles of Eq.(\ref{shiki22}) could be regarded 
as a unique solution with a small 
range coming from the experimental errors and the theoretical uncertainties
in the hadronic matrix elements.

According to the solution of $V^R$ in Eq.(\ref{shiki23}), 
the quantities dealt with in this section acquire the following values,
\begin{eqnarray}
\left| \frac{V^R_{ub}}{V^R_{cb}} \right| &=& 0.093,\qquad \qquad \qquad
\Delta m_{B_d}=3.43\times \left( 1 \quad {{+\,0.78}\atop -\,0.56} \right) 
                  \times 10^{-10} \, \mbox{MeV}, \nonumber \\
\Delta m_K &=& \Delta m^{LL}_K(c,c)+\Delta m^{LR}_K(c,c)+\Delta m^{RR}_K(t,t) 
               \nonumber \\
  &=& (\frac13-1.5)(2.22-0.72+2.02)\times 10^{-12}  
  = (\frac13 -1.5)\times 3.52 \times 10^{-12} \, \mbox{MeV}, \nonumber \\
\varepsilon_K &=& 2.27\times 10^{-3}, \qquad \qquad
  \frac{{\rm Br}(B^-\rightarrow \psi \pi^-)}{{\rm Br}(B^-\rightarrow \psi 
K^-)}=0.053,\label{shiki24}
\end{eqnarray}
where the large errors in $\Delta m_{Bd}$ come from the large uncertainty
in $f_B \sqrt{B_B}=(0.15\pm0.05)$\,GeV.

\vskip 1truecm
\leftline{\bf3.\ \ CP violation }
\vskip 0.4truecm
\noindent
In this section we study "direct" CP-violating parameter $\varepsilon'$ 
in the neutral kaon system and the electric dipole moment of neutron
in our model. \ In addition, we analyze CP violation in the nonleptonic
decay of $B_d$ and $B_s$ mesons into CP-eigenstates, comparing with that
in the Standard Model.

"Direct" CP-violating parameter $\varepsilon'$ arising from the decay
dynamics in neutral kaon system is expressed as follows,
\begin{equation}
\varepsilon'=\frac{1}{\sqrt2} e^{i(\frac{\pi}{2}+\delta_2-\delta_0)}
\frac{{\rm Re} A_2}{{\rm Re} A_0} (\xi_2-\xi_0),
\label{shiki25}
\end{equation}
where $A_{0,2}$ are the transition amplitudes for $\bar{K}^0\rightarrow 
\pi\pi(I=0,2),\xi_{0,2}\equiv {\rm Im} A_{0,2}/{\rm Re} A_{0,2}$, 
and $\delta_{0,2}$ 
are the strong interaction phase shifts. This quantity $\varepsilon'$
has been measured and there are two controversial results from two groups;
$\varepsilon'/\varepsilon=(2.0\pm0.7)\times 10^{-3}$ from CERN 
\cite{Barr93} and
$\varepsilon'/\varepsilon=(0.74\pm0.52\pm0.29)\times 10^{-3}$ from Fermilab
\cite{Gibbons93}.\  
Although these two results are different from each other by two
standard deviations, both measurements would show that 
$\varepsilon'/\varepsilon$ is at the level of $10^{-4}-10^{-3}$. 

In our model, potential contributions to $\varepsilon'$ are the 
eight diagrams shown in Fig.3. \ Imaginary parts of the amplitudes come from 
the last five diagrams. \ By using the calculational procedure with the
QCD corrections given by Ecker and Grimus \cite{Ecker85}, we obtain the 
following 
magnitudes of each amplitude corresponding to the eight diagrams in Fig.3
by use of $V^L$ in Eq.(\ref{shiki1}) and $V^R$ in Eq.(\ref{shiki2});
${\rm Re} A_0\simeq 5.05\times10^{-5}$ MeV and ${\rm Re} A_2\simeq 3.14\times 
    10^{-5}$ MeV for $W_L$-exchange tree diagram (Fig.3(a)),
$\quad ({\rm Re} A_0, {\rm Im} A_0)\simeq (-0.043\cos(\beta-\alpha),\,
    0.043\sin(\beta-\alpha))\times 10^{-5}$ MeV and $({\rm Re} A_2, {\rm Im}
    A_2)\simeq (-0.023\cos(\beta-\alpha),\,0.023\sin (\beta-\alpha))\times
    10^{-5}$ MeV for $W_R$-exchange tree diagram (Fig.3(b)),
$\quad {\rm Re} A_0\simeq 31.0\times10^{-5}$MeV for $W_L$- 
    loop penguin diagram (Fig.3(c)),
$\quad ({\rm Re} A_0, {\rm Im} A_0)\simeq (-(0.428-0.02\cos\delta) 
    \cos(\beta-~\alpha)-0.026\sin\delta \sin(\beta-\alpha),\, 
   (0.428-0.024\cos\delta)\sin(\beta-\alpha)-0.026 \sin\delta\cos(\beta-
    \alpha))\times 10^{-5}$ MeV for $W_R$- loop penguin diagram (Fig.3(d)),
$\quad ({\rm Re} A_0, {\rm Im} A_0)\simeq (0.007 \cos(\lambda-~\alpha)-
    0.317\cos(\lambda-\alpha+\gamma-\delta)-0.833\cos(\lambda-\alpha+\gamma),\,
    -0.007\sin(\lambda-\alpha)+0.317\sin(\lambda-\alpha+\gamma-\delta)
    +0.833\sin(\lambda-\alpha+\gamma)) \times 10^{-5}$ MeV for $W_L-W_R$
    mixing penguin diagram (Fig.3(e)),
$\quad ({\rm Re} A_0, {\rm Im} A_0)\simeq (-0.005\cos(\lambda-\beta)+
    0.459\cos(\lambda-\beta+\gamma-\delta)-0.028\cos(\lambda-\beta+\gamma),
    \,-0.005\sin(\lambda-\beta)+0.459 \sin(\lambda-\beta+\gamma-\delta)
   -0.028\sin(\lambda-\beta+\gamma)) \times10^{-5}$ MeV for
    $W_R-W_L$ mixing penguin diagram (Fig.3(f)),
$\quad ({\rm Re} A_0, {\rm Im} A_0)\simeq(13.6\cos(\lambda-\alpha),
    \,-13.6\sin(\lambda-\alpha)) \times 10^{-5}$ MeV and 
    $({\rm Re} A_2, {\rm Im} A_2) \simeq (-1.18\cos(\lambda-\alpha),\,
    1.18\sin(\lambda-\alpha))\times10^{-5}$ MeV for $W_L-W_R$ 
    mixing tree diagram (Fig.3(g)),\, 
and $({\rm Re} A_0, {\rm Im} A_0)\simeq (-9.46\cos(\lambda-\beta),\,
    -9.46\sin(\lambda-\beta))\times 10^{-5}$ MeV and $({\rm Re} A_2, {\rm Im} 
    A_2) \simeq (0.82\cos(\lambda-\beta),\,0.82\sin(\lambda-\beta))\times 10^
    {-5}$MeV for $W_R-W_L$ mixing tree diagram (Fig.3(h)),\,
where $\lambda$ is the $W_L-W_R$ mixing phase and we used the magnitudes
of the angles $\theta_{12}, \theta_{13}$ and $\theta_{23}$ in 
Eq.(\ref{shiki22}) and $\zeta_g=0.03$ for the $W_L-W_R$ mixing angle 
\cite{Polak92}.         \ By examining the above 
respective magnitude from the eight diagrams in comparison with the average 
value of the two data, $\varepsilon'/\varepsilon=(1.5\pm0.8)\times10^{-3}$
\cite{Barnett96} ,imaginary parts of the contributions from the 
$W_L-W_R$ and $W_R-W_L$ 
mixing tree diagrams prove to be too large, so that we have to constrain 
the phases, $\alpha, \beta$, and $\lambda$ as $\lambda=\alpha$ and 
$\alpha-\beta=\pi$, where the latter constraint is obtained in combination
with the one that $\Delta m^{LR}_K(c,c)$ in Eq.(\ref{shiki14}) should be 
negative to reproduce the experimental value of $\Delta m_K$, as seen 
in Eq.(\ref{shiki24}).If we use these two constraints and 
$\sin\delta=0.0345(\cos\delta\cong-1.0)$ in Eq.(\ref{shiki22}) 
and sum up all of the amplitudes from the eight diagrams,
we obtain the following magnitudes for the real and imaginary parts of $A_0$
and $A_2$,
\begin{eqnarray}
{\rm Re} A_0 &=& 59.6-0.029\cos\gamma-0.027\sin\gamma, \nonumber \\
{\rm Im} A_0 &=& 0.0009+0.0049\cos\gamma+1.00\sin\gamma, \nonumber \\
{\rm Re} A_2 &=& 1.16, \nonumber  \\
{\rm Im} A_2 &=& 0 .  
\label{shiki26}
\end{eqnarray}
In order to reproduce the sign and the order of magnitude of
$\varepsilon'/\varepsilon, \gamma=\pi$ is required for the phase $\gamma$,
as can be seen from ${\rm Im} A_0$ in Eq.(\ref{shiki26}), and we get by 
substituting Eq.(\ref{shiki26}) into Eq.(\ref{shiki25}) as
\begin{equation}
\varepsilon'\simeq e^{i\pi/4}\times 0.93\times 10^{-6},
\label{shiki27}
\end{equation}
where we used $\delta_0-\delta_2 \simeq \pi/4$ \cite{Barnett96}. \ If we use 
the experimental value of $\varepsilon,\  \varepsilon \simeq e^{i\pi/4} \times 
2.28\times 10^{-3}$, we eventually obtain 
\begin{equation}
\varepsilon'/\varepsilon\simeq 0.41\times 10^{-3}. 
\label{shiki28}
\end{equation}   
This value is consistent with the measured value from Fermilab 
\cite{Gibbons93},
and is compatible with the average value of the data from the two
groups, $\varepsilon'/\varepsilon=(1.5\pm0.8)\times10^{-3}\,$ 
\cite{Barnett96}.

Next, we study the electric dipole moment of neutron in our 
model. \ The dipole moment arises from the one loop $W_L-W_R$ mixing
diagrams for the moments of $u$- and $d$- quarks in the 
$SU(2)_L \times SU(2)_R \times U(1)$ model \cite{Ecker83}. \ We neglect here 
the contributions from Higgs loop and exchange diagrams in order to see 
the order of magnitude of the dipole moment. \ The diagrams contributing to 
the moments of $u$ and $d$ quarks ($d_u$ and $d_d$) are those depicted
in Figs.4(a) and (b), respectively.

$d_u$ and $d_d$ are expressed as follows \cite{Ecker83},
\begin{eqnarray}
d_u &=& \frac{4\sqrt2eG_F\zeta_g}{32\pi^2}
        \sum_{j=d,s}m_j {\rm Im}(e^{i\lambda}V^L_{uj}V^{R*}_{uj}) \nonumber \\
     &\times & \left\{ I_1(r_j,s_u)+\frac13 I_2(r_j,s_u)-
        \beta \left[I_1(r_j\beta, s_u\beta) +\frac13I_2(r_j\beta, s_u\beta)
                           \right] \right\},   
\label{shiki29}
\end{eqnarray}
\begin{eqnarray}
d_d &=& \frac{4\sqrt2eG_F\zeta_g}{32\pi^2}
        \sum_{j=u,c}m_j {\rm Im}(e^{i\lambda}V^L_{jd}V^{R*}_{jd}) \nonumber \\
     &\times & \left\{ I_1(r_j,s_d)+\frac23 I_2(r_j,s_d)-
        \beta \left[I_1(r_j\beta, s_d\beta) +\frac23I_2(r_j\beta, s_d\beta)
                           \right] \right\}, 
\label{shiki30}
\end{eqnarray}
where $r_j\equiv(m_j/M_L)^2,\  s_u\equiv(m_u/M_L)^2,\, 
s_d\equiv(m_d/M_L)^2,\, \beta\equiv(M_L/M_R)^2,\, \zeta_g$ and $\lambda$ 
are $W_L-W_R$ mixing angle and phase, respectively, and
\begin{eqnarray}
I_1(r,s) &\simeq& \frac{2}{(1-r)^2}\left(1-\frac{11}{4}r+\frac14r^2
                 -\frac{3r^2\ln r}{2(1-r)}\right), \nonumber    \\
I_2(r,s) &\simeq& \frac{2}{(1-r)^2}\left(1+\frac{1}{4}r+\frac14r^2
                 +\frac{3r\ln r}{2(1-r)}\right).                
\label{shiki31}
\end{eqnarray}
When we use $m_d=10$MeV,\, $m_s=200$MeV,\, $m_u=$5MeV,\, $m_c=$1.5GeV,\,
$\zeta_g$=0.03,\, $M_R$=750GeV and $V^L$ in Eq.(\ref{shiki1}) and $V^R$ 
in Eq.(\ref{shiki23}),we obtain the following values,
\begin{equation}
d_u \ \simeq \  -10.5\sin(\alpha-\lambda) \times 10^{-25}\, {\rm e\cdot cm}, \\
\hspace{6cm}
\label{shiki32}
\end{equation}
\begin{eqnarray}
d_d &\simeq& \left\{ -20.3\sin(\alpha-\lambda)+44.7\sin(\lambda-\alpha
             +\gamma-\delta) +117.4\sin(\lambda-\alpha+\gamma) \right\}
                                              \nonumber             \\
    &\times&    10^{-25}\, {\rm e\cdot cm}.     
\label{shiki33}  
\end{eqnarray} 
We take the $SU(6)$ wave function to calculate the neutron electric dipole
moment to obtain
\begin{equation}
d_n=\frac13(4d_d-d_u)\simeq(96.9\sin\gamma-2.06\cos\gamma)
    \times10^{-25} {\rm e\cdot cm},
\label{shiki34}
\end{equation} 
where we used the phase relation $\lambda=\alpha$, obtained for 
$\varepsilon'/\varepsilon$, and $\sin\delta=0.0345(\cos\delta\cong-1.0)$.
\,If we adopt another relation $\gamma=\pi$, also obtained for
$\varepsilon'/\varepsilon$, we get
\begin{equation}
d_n\simeq 2.1\times 10^{-25}\, {\rm e\cdot cm}.
\label{shiki35}
\end{equation}
This value is larger by a factor 2 than the measured upper limit,
$|d_n|<1.1\times 10^{-25} {\rm e\cdot cm}$ \, \cite{Barnett96}, though it 
is comparable with the limit.

If we relax the relation $\gamma=\pi$ to $\gamma=\pi+x$ with a small
$x$=(0.010,\, 0.012,\, 0.014,\, 0.016),
both the theoretical values of $\varepsilon'/\varepsilon$ and $d_n$
simultaneously vary as
\begin{equation} 
\varepsilon'/\varepsilon = (1.4,\ 1.6,\ 1.8,\ 2.0) \times 10^{-3}\\,
\hspace{2cm}
\label{shiki36}
\end{equation}
\begin{equation}
d_n = (1.1,\ 0.90,\ 0.70,\ 0.50)\times10^{-25}\,{\rm e\cdot cm}.
\label{shiki37}
\end{equation}
This shows that there is a phase value of $\gamma$ around $\gamma\simeq\pi
+0.012$ which satisfies both the experimental value of $\varepsilon'/
\varepsilon$ and the upper limit of $d_n$. \ Therefore, our model agrees  with 
the present experimental situation of $\varepsilon'/\varepsilon$ and $d_n$.

Finally, we discuss CP violation in the nonleptonic decays of $B_d$ and $B_s$
mesons into CP-eigenstates. Integrated CP asymmetry into the CP-eigenstates
is defined by \cite{Carter81}
\begin{equation}
C_f = \frac{{\mit\Gamma}(B^0\rightarrow f)- {\mit\Gamma}(\bar{B}^0\rightarrow f)} 
             {{\mit\Gamma}(B^0\rightarrow f)+{\mit\Gamma} (\bar{B}^0\rightarrow f)}
            =-\frac{x}{1+x^2} \sin\varphi_f,                            \\
\label{shiki38}
\end{equation}
\begin{equation}
\varphi_f = CP(f)(\arg M^*_{12}+2\arg A),
\hspace{3cm}
\label{shiki39}
\end{equation}
where $\mit\Gamma(B^0\rightarrow f)$ is the time-integrated decay rate of 
time-evolved $B^0(t=0)$ into the final hadronic state $(f),\  \bar{f}$ 
the CP-conjugated state of $f,\  x$ the mixing parameter given by
$x=\Delta m_B/\mit\Gamma_B$, CP($f$) \ CP-parity of state $f$,\  $M_{12}$ 
the off diagonal element of mass matrix of the neutral $B^0-\bar{B}^0$
system and $A$ is the weak amplitude of $\bar{B}^0\rightarrow f$ \,decay.
 
In the case of $B_d$ decay, the dominant contribution to $M_{12}$ is
the $W_L-W_R$ box diagram with $c$- and $t$- quark exchange as calculated
in $\S$2, and this has the phase $e^{-i(\alpha-\gamma+\delta)}$.
In Table I, we list the CP asymmetry of various $B_d$ decay modes by 
showing the angle $\varphi_f$ for each weak quark subprocess for the 
case of using phase relations $\gamma=\pi$ and $\beta=\alpha-\pi$
obtained from the study of $\varepsilon'/\varepsilon$ and for the case
without the relations. On the fifth column of Table I, we record 
the predictions of $\varphi_f$ from the Standard Model, where
$\alpha, \beta$ and $\gamma$ are three angles of the unitarity 
triangle, not our phases in $V^R$. \ From Table I, we can see the
remarkable difference between our model and the Standard Model.

For $B_s$ decays, first we estimate the mixing parameter $x_s$ of
$B_s-\bar{B}_s$ mixing. The dominant contribution is the diagram 
in Fig.5. \ The off-diagonal element $M_{12}$ in the $B_s-\bar{B}_s$
system is given by 
\begin{eqnarray}
M_{12} &=& <B_s|H^{LR}_{eff}(c,t)|\bar{B_s}>         \nonumber\\
&=& \frac{2G_F^2 M^2_L \beta_g}{\pi^2}
   \left[ {\left(\frac{m_{Bs}}{m_b+m_s} \right)}^2+\frac16 \right]
   \frac14 f^2_{B_s} B_{B_s} m_{B_s} 
   V^L_{tb} V^{L*}_{cs} V^{R*}_{ts} V^{R}_{cb}       \nonumber\\
&\times& B(x_c,x_t,\beta_g).
\label{shiki40}
\end{eqnarray}
If we take $m_{B_s}$=5.37GeV, \, $m_s$=0.2GeV and $f_{B_s}\sqrt{B_{B_s}}$=
0.15GeV in Eq.(\ref{shiki40}), we get 
\begin{equation}
M_{12}\simeq 43.5e^{-i(\beta-\gamma+\delta)}\times 10^{-10}\,\mbox{MeV},
\label{shiki41}
\end{equation}
and the mixing parameter $x_s$ is obtained to be
\begin{equation}
x_s=\tau_{B_s}\Delta m_{B_s}=2\tau_{B_s} |M_{12}| \simeq 20\, ,
\label{shiki42}
\end{equation}
where we used $\tau_{B_s}\simeq 1.5 \times10^{-12}$ s.\ 
When we use $f_{B_s}\sqrt{B_{B_s}}=(0.15\pm0.05)$GeV,
we obtain $x_s=20\times \left( 1 \, {+0.78\atop -0.56}\right)$.
From Eq.(\ref{shiki42}), we expect that $B_s-\bar{B}_s$ system have 
a very rapid 
oscillation as compared with the $B_d-\bar{B}_d$ system. \ This feature
is the same as in the Standard Model. The $x_s$ has recently been 
measured and obtained to be $x_s>9$ \, \cite{BuskulicII96}.

CP asymmetry $C_f$ in Eq.(\ref{shiki38}) for $B_s$ meson decays into hadronic 
CP-eigenstates is below 0.05 due to the large oscillation $x_s\sim20$,
and this is about 1/10 the CP asymmetry in $B_d$ decays. \ Since the phase
of $M_{12}$ for $B_s$ decay is the same as for the $B_d$ decay up to minus 
sign as can be seen from Eq.(\ref{shiki41}), which is obtained by use 
of the relation
$\alpha-\beta=\pi$, the structure of the angle $\varphi_f$ of $C_f$ is 
same between $B_d$ and $B_s$ decays up to the additional $\pi$ in $B_d$ 
decays, as seen in Table II. \ This point is remarkably different from the
Standard Model. \ The typical predictions about the CP asymmetry from our
model are the following,
\begin{eqnarray}
C_f(B_d\rightarrow D^+D^-) &=& C_f(B_d\rightarrow D^0_1K_s)
                              =C_f(B_d\rightarrow \pi^+\pi^-) \nonumber \\
   &=&-C_f(B_d\rightarrow D^0_1\pi^0)\simeq -C_f(B_d\rightarrow \psi K_s),
\label{shiki43}
\end{eqnarray} 
\begin{eqnarray}
C_f(B_s\rightarrow \psi K_s) &=& C_f(B_s\rightarrow D^0_1\phi)
                              =C_f(B_s\rightarrow \rho^0K_s) \nonumber \\
   &=&-C_f(B_s\rightarrow D^0_1K_s)\simeq -C_f(B_s\rightarrow D^+_sD^-_s),
\label{shiki44}
\end{eqnarray}
for the case of using $\gamma=\pi$. The last equalities $\simeq$ of both  
Eqs.(\ref{shiki43}) and (\ref{shiki44}) come from the contamination of 
small phase $\delta (\sim2^{\circ})$.
 
As can be seen from the following relation
\begin{equation}
\frac12 \left[ \varphi(B_d\rightarrow \psi K_s)
               +\varphi(B_d\rightarrow \pi^+\pi^-)
               +\varphi(B_s\rightarrow \rho^0 K_s) \right]
              =\frac12(\alpha+\delta)\ne\pi,
\label{shiki45}
\end{equation}
the angles of these three modes do not necessarily construct the unitarity
triangle in our model. \ This fact generally holds in the 
$SU(2)_L \times SU(2)_R \times U(1)$ model in which a new particle ($W_R$)
causes the $B_{d(s)}$ decays \cite{Kurimoto96}.
\vskip 1truecm

\leftline{\bf 4.\  Discussions and conclusions}
\vskip 0.4truecm
\noindent
Hadronic decays of $B$ mesons caused by the $b$ to $c$ transition 
originate from the quark subprocesses $b\rightarrow c\bar{c}s, \,
b\rightarrow c\bar{u}d$ and $b\rightarrow c\bar{u}s$,
in addition to the process $b\rightarrow c\bar{c}d$ which was used to
constrain the mixing angles of $V^R$ by use of the mode $B^-\rightarrow 
\psi\pi^-$
in \S 2. \ As for the process  $b\rightarrow c\bar{c}s$, the decay rate of
$B_d\rightarrow D^{*-}D^+_s$ is related to the differential decay rate
of $B_d \rightarrow D^{*-}l^+\nu$ at the momentum transfer squared
$q^2=m^2_{D_s}$ to the lepton pair by the following equation by using the 
factorization hypothesis and the heavy quark effective theory 
\cite{Bjorken89,Bortoletto90},
\begin{equation}
\frac{\mit\Gamma(B_d\rightarrow D^{*-}D^+_s)}
     {d\mit\Gamma(B_d\rightarrow D^{*-}l^+\nu)/dq^2|_{q^2=m^2_{D_s}}}
  =6\pi^2f^2_{D_s}|V_{cs}|^2,
\label{shiki46}
\end{equation}
where $f_{D_s}$ is the decay constant of $D_s$ meson. 
\ By using the experimental
values $f_{D_s}=430\ {+150+40\atop -130-40}$\,MeV \cite{Bai95},
${\rm Br}(B_d\rightarrow D^{*-}D^+_s)=(1.2\pm0.6)\times 10^{-2}$\,
\cite{Barnett96} 
and $d{\rm Br}(B_d\rightarrow D^{*-}l^+\nu)/dq^2|_{q^2=m^2_{D_s}}=(0.47\pm0.08)
\times10^{-2}\mbox{GeV}^{-2}$\,\cite{Bortoletto90}, we obtain 
$|V^R_{cs}|=0.23-0.96$ from Eq.(\ref{shiki46}),
since $b\rightarrow c\bar{c}s$ is mediated by $W_R$ exchange in our model.
This value of $|V^R_{cs}|$ agrees with the one in Eq.(\ref{shiki23}).

As for the process $b\rightarrow c\bar{u}d$, decay rate of the relevant
mode $B_d\rightarrow D^{*-}\pi^+$ and the different decay rate of 
$B_d\rightarrow D^{*-}l^+\nu$ at $q^2=m^2_{\pi}$ are related by the 
following equation in the same way as for the above process,
\begin{equation}
\frac{\mit\Gamma(B_d\rightarrow D^{*-}\pi^+)}
     {d\mit\Gamma(B_d\rightarrow D^{*-}l^+\nu)/dq^2|_{q^2=m^2_{\pi}}}
  =6\pi^2f^2_{\pi}|V_{ud}|^2.
\label{shiki47}
\end{equation}
When we use the experimental values ${\rm Br}(B_d\rightarrow D^{*-}\pi^+)
=(2.6\pm0.4)\times10^{-3}$\,\cite{Barnett96} and
$d{\rm Br}(B_d\rightarrow D^{*-}l^+\nu)/dq^2|_{q^2=m^2_{\pi}}=(0.25\pm0.08)
\times10^{-2}\mbox{GeV}^{-2}$\, \cite{Bortoletto90} and $f_\pi=0.132$GeV, 
we obtain
$|V^R_{ud}|=0.80-1.0$ in our model and this is satisfied by the value
$|V^R_{ud}|=0.984$ in Eq.(\ref{shiki23}).

As for the final process $b\rightarrow c\bar{u}s$, decay rates of the modes
coming from this process like $B_d\rightarrow D^{*-}K^+$ have not been
measured. \ In our model the branching ratio of $B_d\rightarrow D^{*-}K^+$
is related to the observed ratio of ${\rm Br}(B_d\rightarrow D^{*-}\pi^+)
=(2.6\pm0.4)\times10^{-3}$ by the following equation,
\begin{equation}
\frac{{\rm Br}(B_d\rightarrow D^{*-}K^+)}{{\rm Br}(B_d\rightarrow D^{*-}\pi^+)}
\approx \frac{|V^R_{us}|^2}{|V^R_{ud}|^2}\,.
\label{shiki48}
\end{equation}
We can predict as ${\rm Br}(B_d\rightarrow D^{*-}K^+)=(6.1\pm1.0)\times10^{-5}$
by using the magnitudes of $|V^R_{us}|$ and $|V^R_{ud}|$ in Eq.(\ref{shiki23}).
\ This value is about 1/2 of the predicted value $(1.3\pm0.2)\times10^{-4}$
of the Standard Model.

In conclusion, we investigated the purely right-handed $b$ to $c$ and
$b$ to $u$ coupling model with full mixing angles and phases in $V^R$
in the framework of $SU(2)_L\times SU(2)_R \times U(1)$ gauge model.
\ The model has turned out to be viable still under the new circumstance
that the right-handed $W$ boson mass $M_R>720$\,GeV. \ It satisfies the 
experimental values of CP-violating parameter $\varepsilon'/\varepsilon$
in the neutral kaon system and the upper limit of neutron electric dipole 
moment.\ The model would predict a remarkably different pattern of CP 
asymmetry in $B_d$ and $B_s$ decays into hadronic CP-eigenstates from that in 
Standard Model. We expect that the asymmetry will be measured to test the 
model at the $B$- factories under construction.
\vskip 1truecm 

\centerline{ACKNOWLEDGEMENTS}
\vskip 0.4truecm

The author would like to express his gratitude to T. Hattori, Z. Hioki
and T. Hasuike for useful discussions and warm hospitality at their
Institute to complete this work. \ He is grateful to S. Wakaizumi for 
helpful discussions and careful reading of the manuscript. \ He also thanks
to T. Kurimoto for providing a suggestion on the number of phases in
the left- and right-handed mixing matrices.

\newpage
\begin{table}[p]
\caption{List of the angles $\varphi_f$ in the CP asymmetry $C_f$ for 
$B_d$ decays into hadronic CP-eigenstates without and with the phase 
relations $\gamma=\pi$ and $\beta=\alpha-\pi$ for various quark subprocesses 
in our model and in the Standard Model(fifth column).}
\label{table1}
\begin{center}
\begin{tabular}{llccc}\hline\hline 
  {Subprocess}&{$B_d$ decay mode}&{$\varphi_f$}
              &{$\varphi_f$with{$\gamma=\pi,\atop \beta=\alpha-\pi$}}
                            &{$\varphi_f$(S.M.)} \\  \hline
$b\rightarrow c\bar{c}s$     &  $B_d\rightarrow \psi K_s$    &
    $\alpha-\gamma+\delta$     &  $\pi+\alpha+\delta$        &  $2\beta$   \\
$b\rightarrow c\bar{u}d$     &  $B_d\rightarrow D^0_1\pi^0$  &
    $\alpha+\gamma-\delta$     &  $\pi+\alpha-\delta$        &  $2\beta$   \\
$b\rightarrow c\bar{c}d$     &  $B_d\rightarrow D^+D^-$      &
    $-(\alpha-\gamma-\delta)$  &  $-(\pi+\alpha-\delta)$     & $-2\beta$   \\
$b\rightarrow c\bar{u}s$     &  $B_d\rightarrow D^0_1K_s$    &
    $-(\alpha+\gamma-\delta)$  &  $-(\pi+\alpha-\delta)$     &  $-2\beta$  \\
$b\rightarrow u\bar{u}d$     &  $B_d\rightarrow \pi^+\pi^-$  &
    $-(\alpha-\gamma-\delta)$  &  $-(\pi+\alpha-\delta)$     &  $2\alpha$   \\
$b\rightarrow u\bar{u}s$     &  $B_d\rightarrow \pi^0Ks$     &
    $\alpha-\gamma-\delta$     &  $\pi+\alpha-\delta$        &  $?$       \\ 
$b\rightarrow u\bar{c}s$     &  $B_d\rightarrow D^0_1K_s$    &
    $-(\alpha-3\gamma+\delta)$ &  $-(\pi+\alpha+\delta)$     &  $2\alpha$  \\
$b\rightarrow u\bar{c}d$     &  $B_d\rightarrow D^0_1\pi^0$  &
    $\alpha-3\gamma-\delta$    &  $\pi+\alpha-\delta$        &  $-2\alpha$  \\
$b\rightarrow s\bar{s}s$     &  $B_d\rightarrow \phi K_s$    &
    $\alpha-\gamma+\delta$     &  $\pi+\alpha+\delta$        &  $2\beta$    \\
$b\rightarrow s\bar{d}d$     &  $B_d\rightarrow \pi^0 K_s$   &
    $\alpha-\gamma+\delta$     &  $\pi+\alpha+\delta$        &  $2\beta$    \\
$b\rightarrow d\bar{s}s$     &  $B_d\rightarrow K_s K_s$     &
    $-(\alpha-\gamma+\delta)$  &  $-(\pi+\alpha+\delta)$     &  $-2\beta$  \\
$b\rightarrow d\bar{d}d$     &  $B_d\rightarrow \pi^0\pi^0$  &
    $-(\alpha-\gamma+\delta)$  &  $-(\pi+\alpha+\delta)$     &  $-2\beta$  \\
\hline
\end{tabular}
\end{center}
\end{table}

\begin{table}[p]
\caption{List of the angles $\varphi_f$ in the CP asymmetry $C_f$ for 
$B_s$ decays into hadronic CP-eigenstates without and with the phase 
relations $\gamma=\pi$ and $\beta=\alpha-\pi$ for various quark subprocesses 
in our model and in the Standard Model(fifth column).}
\label{table2}
\begin{center}
\begin{tabular}{llccc}\hline\hline 
{Subprocess}&{$B_s$ decay mode}&{$\varphi_f$}
              &{$\varphi_f$with{$\gamma=\pi,\atop \beta=\alpha-\pi$}}
                            &{$\varphi_f$(S.M.)} \\  \hline
$b\rightarrow c\bar{c}s$     &  $B_s\rightarrow D^+_sD^-_s$    &
    $\beta+\gamma-\delta-2\alpha$     &  $-(\alpha+\delta)$        &  $0$   \\
$b\rightarrow c\bar{u}d$     &  $B_s\rightarrow D^0_1K_s$      &
    $\beta-\gamma+\delta-2\alpha$     &  $-(\alpha-\delta)$        &  $0$   \\
$b\rightarrow c\bar{c}d$     &  $B_s\rightarrow \psi K_s$       &
    $-(\beta+\gamma+\delta-2\alpha)$  &  $\alpha-\delta$           &  $0$   \\
$b\rightarrow c\bar{u}s$     &  $B_s\rightarrow D^0_1\phi$     &
    $-(\beta-\gamma+\delta-2\alpha)$  &  $\alpha-\delta$           &  $0$   \\
$b\rightarrow u\bar{u}d$     &  $B_s\rightarrow \rho^0K_s$     &
    $-(\beta+\gamma+\delta-2\alpha)$  &  $\alpha-\delta$       &  $2\gamma$  \\
$b\rightarrow u\bar{u}s$     &  $B_s\rightarrow K^+K^-$        &
    $\beta+\gamma+\delta-2\alpha$  &  $-(\alpha-\delta)$       &  $?$  \\
$b\rightarrow u\bar{c}s$     &  $B_s\rightarrow D^0_1\phi$     &
    $-(\beta+3\gamma-\delta-2\alpha)$  &  $\alpha+\delta$      &  $2\gamma$  \\
$b\rightarrow u\bar{c}d$     &  $B_s\rightarrow D^0_1K_s$      &
    $\beta+3\gamma+\delta-2\alpha$     &  $-(\alpha-\delta)$   &  $-2\gamma$ \\
$b\rightarrow s\bar{s}s$     &  $B_s\rightarrow \acute{\eta}\acute{\eta}$   &
    $\beta+\gamma-\delta-2\alpha$     &  $-(\alpha+\delta)$    &  $0$        \\
$b\rightarrow s\bar{d}d$     &  $B_s\rightarrow K_sK_s$        &
    $\beta+\gamma-\delta-2\alpha$     &  $-(\alpha+\delta)$    &  $0$        \\
$b\rightarrow d\bar{s}s$     &  $B_s\rightarrow \phi K_s$      &
    $-(\beta+\gamma-\delta-2\alpha)$     &  $\alpha+\delta$    &  $2\beta$   \\
$b\rightarrow d\bar{d}d$     &  $B_s\rightarrow \pi^0 K_s$     &
    $-(\beta+\gamma-\delta-2\alpha)$     &  $\alpha+\delta$    &  $2\beta$   \\
\hline
\end{tabular}
\end{center}
\end{table}

\newpage

\centerline{\bf Figure captions}
\vskip 0.5truecm
\noindent
{\bf Fig.1.} The $W_L-W_R$ box diagram with $c$- and $t$- quark exchanges
for calculating $\Delta m_{B_d}$.

\vskip 0.3truecm
\noindent
{\bf Fig.2.} The box diagrams for calculating $\Delta m_K$ ; (a)$W_L-W_L$
box diagram with two $c$- quark exchanges, (b)$W_L-W_R$ box with two $c$-
quark exchanges, and (c)$W_R-W_R$ box with two $t$- quark exchanges.

\vskip 0.3truecm
\noindent
{\bf Fig.3.} The diagrams for calculating CP-violating parameter 
$\varepsilon'$ ; (a)$W_L$-exchange tree diagram, (b)$W_R$-exchange
tree diagram, (c)$W_L$-loop penguin diagram, (d)$W_R$-loop penguin diagram,
 (e)$W_L-W_R$ mixing penguin diagram, (f)$W_R-W_L$ mixing penguin diagram,
 (g)$W_L-W_R$ mixing tree diagram, (h)$W_R-W_L$ mixing tree diagram.

\skip 0.3truecm
\noindent
{\bf Fig.4.} The one loop diagrams for calculating (a) $u$- and (b) $d$- quark
electric dipole moment.

\vskip 0.3truecm
\noindent
{\bf Fig.5.} The $W_L-W_R$ box diagram with $c$- and $t$- quark exchanges for
calculating $B_s-\bar{B}_s$ mixing.

\end{document}